\documentclass[aps,prl,superscriptaddress,floatfix,showpacs,notitlepage,twocolumn]{revtex4-1}
\usepackage[latin9]{inputenc}
\usepackage{color}
\usepackage{textcomp}
\usepackage{amsmath}
\usepackage{amssymb}
\usepackage{mathtools}
\usepackage{graphicx}
\usepackage{esint}
\usepackage{framed} 
\usepackage{xcolor}
\usepackage{hyperref}
\hypersetup{
  linkcolor = black,
  citecolor  = blue,
  urlcolor = black,
  colorlinks = true,
}
\makeatletter

\usepackage{amsfonts}
\usepackage{bbold}
\usepackage{natbib}
\usepackage{svg}
\usepackage{float}

\pdfoutput=1
\newcommand{\ket}[1]{|#1 \rangle}
\newcommand{\bra}[1]{\langle#1 |}
\newcommand{\braket}[2]{\langle #1 | #2 \rangle}

\newcommand{\ii}{\mathrm{i}}
\newcommand{\ee}{\mathrm{e}}

\begin{document}

\title{Squeezing Quantum Many-Body Scars}

\author{Bennet Windt}
\address{Institute for Quantum Optics and Quantum Information, Austrian Academy of Science, 6020 Innsbruck, Austria }
\address{Blackett Laboratory, Imperial College London, Prince Consort Road, London SW7 2AZ, UK }
\author{Hannes Pichler}
\address{Institute for Quantum Optics and Quantum Information, Austrian Academy of Science, 6020 Innsbruck, Austria }
\address{Institute for Theoretical Physics, University of Innsbruck, 6020 Innsbruck, Austria }

\begin{abstract}
We develop an analytical  approach for the description of quantum many-body scars in PXP models. We show that the scarred dynamics in the PXP model on a complete bipartite graph can be interpreted as a one-dimensional chiral scattering problem, and solve this problem analytically. The insights from this analysis allow us to predict that dynamical signatures of scars in PXP models can be enhanced by spin squeezing the initial states. We show numerically that this stabilization mechanism applies not only to the complete bipartite graph but also to one- and two-dimensional lattices, which are relevant for Rydberg atom array experiments. Moreover, our findings provide a physical motivation for Hamiltonian deformations reminiscent of those known to produce perfect scars.
\end{abstract}

\date{\today}

\maketitle

When a quantum many-body system is brought out of equilibrium, its constituents  typically relax to their individual equilibrium states, in a process referred to as thermalization \cite{Deutsch:1991ju,Srednicki:1994dl}. Importantly, thermalization occurs even in closed quantum systems, since the different constituents of an interacting many-body system can act as a reservoir for each other. This paradigm provides a very powerful framework for understanding the emergence of statistical mechanics from a microscopic perspective \cite{{Rigol:2008bf}}. 

Recently, quantum many-body scarring, a novel phenomenon that defies this paradigm, has gained significant interest \cite{serbynQuantumManybodyScars2021}. Quantum many-body scars (QMBS) have been discovered in quantum simulation experiments with Rydberg atom arrays \cite{Bernien:2017bp}, where  certain ordered initial states undergo periodic dynamics, in contrast to the expected relaxation of local observables to stationary, thermal values.
Since then, significant research efforts have been devoted to uncovering mechanisms which underlie and enhance this phenomenon \cite{khemaniSignaturesIntegrabilityDynamics2019,choiEmergentSUDynamics2019,bluvsteinControllingQuantumManybody2021,michailidisStabilizingTwodimensionalQuantum2020}.
While scars have since been discovered also in other models~\cite{okTopologicalManybodyScar2019,markEnsuremathEtaPairing2020,moudgalyaQuantumManybodyScars2020,markUnifiedStructureExact2020,bullSystematicConstructionScarred2019,linExactQuantumManyBody2019,choiEmergentSUDynamics2019,schecterWeakErgodicityBreaking2019,iadecolaQuantumManybodyScars2019,chattopadhyayQuantumManybodyScars2020,moudgalyaLargeClassesQuantum2020}, their origin in the original Rydberg model remains not completely understood. Various methods to address this problem have been employed, including Krylov subspace methods \cite{Turner:2018iz,turnerQuantumScarredEigenstates2018} and variational approaches \cite{Ho:2019gv,michailidisSlowQuantumThermalization2020}. 

In this work we introduce a complementary approach, by mapping the dynamics associated with QMBS in Rydberg atom arrays onto a scattering problem in an appropriate limiting case. We provide an analytical solution of this scattering problem, whose interpretation offers a new perspective on the mechanisms underlying QMBS. As an immediate application we also identify physical mechanisms by which the dynamical signatures of scars can be enhanced. Specifically, the scattering matrix suggests an enhancement of the periodic revivals by squeezing the initial states, which we confirm numerically for various lattices. Moreover, our approach also provides a clear physical motivation for Hamiltonian deformations to PXP models that result in perfect QMBS, which have been introduced ad hoc in previous studies \cite{khemaniSignaturesIntegrabilityDynamics2019,choiEmergentSUDynamics2019}.  

\textit{Model.---} To understand the dynamics of scars in Rydberg atom arrays, we work within the well-established PXP approximation \cite{lesanovskyLiquidGroundState2012}. 
Specifically, we consider a  graph $G=(V,E)$ with vertices $V$ and edges $E$, which induces a PXP model as follows: We identify each vertex $i\in V$ with a qubit with states $\ket{0}_i$ and $\ket{1}_i$, and introduce a projection operator $P_e$ for each edge $e=\{i,j\}\in E$ as $P_e=\mathbb{1}-\ket{1}_i\bra{1}\otimes \ket{1}_j\bra{1}$.  We call the common eigenspace of all these commuting projectors with eigenvalue $+1$ the constrained space $\mathcal{H}_{P}$. It contains all states with no neighboring qubits simultaneously in state $\ket{1}$, akin to a nearest-neighbour approximation of the Rydberg-blockade constraint~\cite{jakschFastQuantumGates2000a,saffmanQuantumInformationRydberg2010,weimerRydbergQuantumSimulator2010a}. The projector onto this space is denoted $\mathcal{P}=\prod_{e\in E}P_e$.
Within $\mathcal{H}_P$, dynamics are governed by the so-called PXP Hamiltonian, 
\begin{align}
    H=\frac{\Omega}{2}\sum_{i=1}^N\mathcal{P}\sigma_i^x\mathcal{P},
\end{align}
which has a simple interpretation: each qubit undergoes single particle Rabi oscillations if all its neighbors are in the state $\ket{0}$, while its dynamics are frozen if at least one of its neighbors is in the state $\ket{1}$. Despite its simple form, the conditional dynamics render the Hamiltonian generically non-integrable \cite{serbynQuantumManybodyScars2021}.

\begin{figure*}[t]
\includegraphics[width=\textwidth]{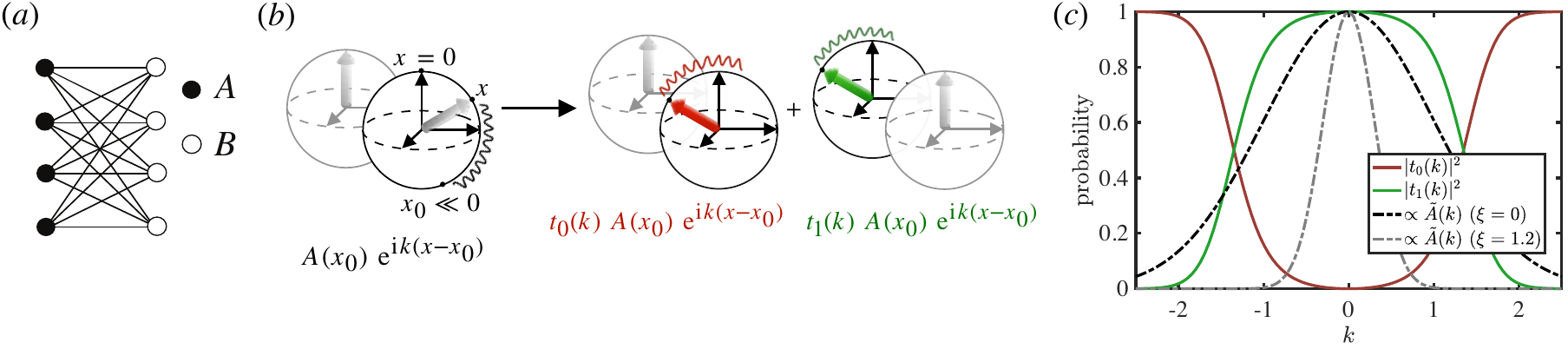}
  \caption{(a) Example of a CBG on $N=8$ sites. (b) Schematic illustration of the scattering problem. States on either bipartition of a CBG are represented on a generalised Bloch sphere, parametrised by a position observable $x$ along a great circle. An incident plane wave (wavenumber $k$) on partition $A$, with partition $B$ in the vacuum state (left), is scattered into a superposition with one of the two partitions in the vacuum state (right). (c) Scattering probabilities $|t_0(k)|^2$ and $|t_1(k)|^2$ and unsqueezed ($\xi=0$) and squeezed ($\xi=1.2$) wavefunctions of the $\mathbb{Z}_2$ state.}
    \label{fig:scattering_problem}
\end{figure*}

We are interested in PXP models on bipartite graphs, $V=A\cup B$, where $A$ and $B$ denote the two partitions. On such graphs, $\mathcal{H}_P$ contains two extremal states $\ket{\mathbb{Z}_2^{A}}$ and $\ket{\mathbb{Z}_2^{B}}$, where all qubits $i\in A$ are in the state $\ket{1}_i$ and all qubits $i\in B$ are in state $\ket{0}_i$ for $\ket{\mathbb{Z}_2^{A}}=\ket{1}^{\otimes i\in A}\ket{0}^{\otimes i\in B}$ and vice versa for $\ket{\mathbb{Z}_2^{B}}=\ket{0}^{\otimes i\in A}\ket{1}^{\otimes i\in B}$.
These states play an important role for scars in PXP models; specifically, it was discovered in Ref.~\cite{Bernien:2017bp} that in one dimension (1D), these states undergo approximate periodic dynamics, with alternating revivals of the two $\mathbb{Z}_2$ states:  $\ee^{-\ii HT_{1D}/2}\ket{\mathbb{Z}_2^A}\approx \ket{\mathbb{Z}_2^B}$, and $\ee^{-\ii HT_{1D}/2}\ket{\mathbb{Z}_2^B}\approx \ket{\mathbb{Z}_2^A}$, with $T_{1D}\approx2\pi\times1.51\Omega^{-1}$. Recently, experiments with Rydberg atom arrays observed analogous scars in bipartite lattices in two dimensions (2D)~\cite{bluvsteinControllingQuantumManybody2021}. Importantly, such periodic dynamics imply that the $\mathbb{Z}_2$ states have non-negligible overlap with only an extensive number of equally spaced, atypical eigenstates of the Hamiltonian, the so-called quantum many-body scars~\cite{Turner:2018iz}.

To develop an intuitive interpretation of these periodic dynamics, we find it convenient to first consider a special limiting case: the complete bipartite graph  (CBG), where each site in $A$ is neighboring to each site in $B$, i.e. $E=\{\{i,j\}~|~i\in A,~j\in B\}$, with $|A|=|B|=N/2$, resulting in $\mathrm{dim}(\mathcal{H}_P)=2^{N/2+1}-1$ (see Fig.~\ref{fig:scattering_problem}). In the thermodynamic limit $N\rightarrow\infty$, we will show below that the scar dynamics on this graph map onto a 1D scattering problem that can be solved analytically. 

On the CBG, the PXP Hamiltonian simplifies to
\begin{align}\label{eq:HCBG}
    H=\hat{S}_A^x\otimes\ket{\phi}_B\bra{\phi}+\ket{\phi}_A\bra{\phi}\otimes\hat{S}_B^{x}\,,
\end{align}
where we have set $\Omega=1$  for simplicity and defined global spin operators in bipartiton $A$ as $\hat{S}^\mu_{A}=1/2\sum_{i\in A}\sigma^\mu_{i}$ ($\mu\in \{x,y,z\}$) and analogously for $B$.  The states $\ket{\phi}_A=\ket{0}^{\otimes i\in A}$ and $\ket{\phi}_B=\ket{0}^{\otimes i\in B}$ denote the state where all qubits $i\in A$ or $i\in B$ are in the $\ket{0}_i$ state, respectively. We refer to $\ket{\phi}$ as the vacuum state. Whenever it is clear from the context we drop the subscripts indicating the partition. For the CBG, a general time-dependent state may be written as
\begin{align}
    \label{eq:bipartite state}
    \ket{\psi(t)}=\ket{A(t)}\otimes\ket{\phi}+\ket{\phi}\otimes\ket{B(t)}\,.
\end{align}
Note that $\ket{\psi(t)}$ is normalised but $\ket{A(t)}$ and $\ket{B(t)}$ are not individually normalised. This form highlights explicitly the fact that partition $A$ can only be in a state other than the vacuum state if partition $B$ is in the vacuum state, and vice versa. The time-dependent Schr{\"o}dinger equation for a state~\eqref{eq:bipartite state} associated with the Hamiltonian~\eqref{eq:HCBG} takes the form of two coupled equations:
\begin{equation}
\begin{aligned}
\label{eq:TDSE}
    \ii\partial_t\ket{A(t)} &=\hat{S}^x\ket{A(t)}+\langle\phi|B(t)\rangle ~\hat{S}^x\ket{\phi}\\
    \ii\partial_t\ket{B(t)} &=\hat{S}^x\ket{B(t)}+\langle\phi|A(t)\rangle ~\hat{S}^x\ket{\phi}\,.
\end{aligned} 
\end{equation}
Evidently, the coupling is governed by the overlap of the state on either partition with the vacuum state. For a random state, this overlap is exponentially small in the system size $N$ and can be neglected. In this approximation, the decoupled equations can be trivially solved: the solution is simply a global rotation of all spins in the active partition, i.e. $\ket{A(t)}=e^{-i\hat S^x t}\ket{A(0)}$ and $\ket{B(t)}=e^{-i\hat S^x t}\ket{B(0)}$, respectively. Importantly, one can check for each approximate solution that the overlap with the vacuum state remains exponentially small at all times, thus justifying the decoupling approximation a posteriori. Crucially, for the $\mathbb{Z}_2$ initial states the decoupling approximation breaks down after some finite time. 
This allows us to distinguish two types of dynamics induced by \eqref{eq:HCBG}: trivial oscillatory dynamics of uncoupled partitions, and dynamics of coupled partitions. Even though the former is largely irrelevant for our analysis, it is important to stress that it is associated with an approximate integrability of each partition. This is a peculiarity of the CBG: on other bipartite graphs, the PXP Hamiltonian is chaotic and thermalizing for generic initial states \cite{Turner:2018iz}. This approximate integrability on the CBG can be easily removed without affecting the dynamics of the $\mathbb{Z}_2$ states, for example, by including additional integrability-breaking terms in the Hamiltonian that vanish on all states that are invariant under permutations of spins within a partition. An explicit example of such a construction is given in Ref.~\cite{choiEmergentSUDynamics2019}.

\textit{Scattering problem. --} For the remainder of this work, we focus on the non-trivial dynamics resulting from the $\mathbb{Z}_2$ initial states. For the sake of concreteness, we choose $A$ as the initially active partition, i.e. $\ket{A(0)}=\ket{S,\theta_0=\pi,\varphi_0=\pi/2}$ and $\ket{B(0)}=0$, where we have defined spin coherent states $\ket{S,\theta,\varphi}\equiv(\cos(\tfrac{\theta}{2})\ket{0}+\ee^{-\ii\varphi}\sin(\tfrac{\theta}{2})\ket{1})^{\otimes N/2}$ (see Supplemental Material [SM] for details). Neglecting any coupling between the partitions, these states evolves as $\ket{A(t)}=\ket{S,\pi-t,\pi/2}$ and $\ket{B(t)}=0$. The overlap of the spin coherent state $\ket{S,\theta,\pi/2}$ with the vacuum state is given by $\braket{\phi}{S,\theta,\pi/2}=\cos^{N/2}(\theta/2)\xrightarrow{N\gg1} \exp(-N\theta^2/8)$~\cite{radcliffe_properties_1971}.
This justifies the decoupling approximation for times $t\lesssim \pi- 2/\sqrt{N}$. At later times, however, the state then enters a narrow neighborhood of the vacuum state and the coupling between the two partitions needs to be taken into account properly. 
A suitable framework to tackle the dynamics in this regime is the Holstein-Primakoff transformation (see SM for details), which embeds the Hilbert space containing states close to the vacuum state in that of a bosonic mode via
\begin{equation}
\begin{aligned}
    \label{eq:HP main text}
    \hat{x}&=\frac{\ii}{\sqrt{2}}(\hat{a}^\dagger-\hat{a})=-\frac{1}{\sqrt{S}}\lim_{S\to\infty}\hat{S}^y\,,\\
    \hat{p}&=\frac{1}{\sqrt{2}}(\hat{a}^\dagger+\hat{a})=\frac{1}{\sqrt{S}}\lim_{S\to\infty}\hat{S}^x\,,
\end{aligned}    
\end{equation}
where $\hat{a}$ is the bosonic ladder operator that annihilates the vacuum $\ket{0}=\lim_{N\to\infty}\ket{\phi}$, and where $S=N/4$. The eigenstates of $\hat x$ and $\hat p$ with eigenvalues $x$ and $p$ are denoted by $\ket{x}$ and $\ket{p}$ respectively. Here, $x$ is identified with the (re-scaled) position on the $\varphi=\pi/2$ great circle of the spin coherent state Bloch sphere, $x\sim \sqrt{S}\theta$.

We proceed in solving the coupled equations \eqref{eq:TDSE} in the vicinity of the vacuum state by constructing the eigenstates of the associated time-independent Sch\"odinger equation in the position basis. That is, we consider solutions of the form $\ket{A(t)}=\ee^{-\ii Et}\ket{A}$ and $\ket{B(t)}=\ee^{-\ii Et}\ket{B}$, where the eigenstates in the position basis, $A(x)=\langle{x}\ket{A}$ and $B(x)=\langle{x}\ket{B}$, satisfy 
\begin{equation}
\begin{aligned}
    \label{eq:differential equations}
    A'(x)&=\ii kA(x)-\langle\phi|B\rangle~\phi'(x)\\
    B'(x)&=\ii kB(x)-\langle\phi|A\rangle~\phi'(x)\,.
\end{aligned}    
\end{equation}
Here, we have introduced the re-scaled energy eigenvalue  $k=E/\sqrt{S}$ and we use the notation $f'(x)$ to indicate the derivative of $f(x)$ with respect to $x$. The vacuum state in this position basis is $ \phi(x)=\pi^{-1/4}\ee^{-x^2/2}$ (see SM).

The coupled differential equations \eqref{eq:differential equations} can be interpreted as a scattering problem for two chiral channels coupled by a short range non-local potential. In both channels, $A$ and $B$, the system can freely propagate towards increasing $x$, except in the vicinity of $x=0$, where the channels are coupled. There the system can either remain in the same channel or scatter into the other channel, before leaving the scattering region.  The relevant data characterizing the scattering solutions are thus the (energy dependent) transmission coefficients $t_0(k)$ and $t_1(k)$, associated with no change of the channel and a flip of the channel in the scattering process, respectively (see Fig.~\ref{fig:scattering_problem}). The chiral nature and the elasticity of the scattering process guarantee $|t_0(k)|^2+|t_1(k)|^2=1$.

We can formally solve the differential equations~\eqref{eq:differential equations} to obtain analytical expressions for both scattering coefficients $t_0(k)$ and $t_1(k)$, conveniently expressed using the short-hand notation for the (iterated) Gaussian integrals
\begin{align}
\begin{split}
    &G(k)=\int_{-\infty}^\infty\phi(z)\ee^{\ii kz}~dz \\
    &H(k,x)=\int_{x_0}^x\phi'(z)\ee^{-\ii kz}~dz \\
    &F(k)=\int_{-\infty}^\infty\phi(z)\ee^{\ii kz}H(k,z)~dz    
\end{split}
\end{align}
The solutions read
\begin{align}
\begin{split}
    t_{0}(k) &=\frac{1+G(k)F(k)H(k)}{1-F(k)^2} \\
    t_{1}(k) &=\frac{G(k)H(k)}{1-F(k)^2}\,,    
\end{split}
\end{align}
with $H(k)=\lim_{x\to\infty}H(k,x)$ (see SM for details).
The transmission coefficients are system size independent and the associated probabilities $|t_0(k)|^2$ and $|t_1(k)|^2$ are shown in Fig.~\ref{fig:scattering_problem}. Importantly, $t_0(0)$ vanishes exactly, corresponding to perfect transfer between the channels at zero energy, i.e. $|t_1(0)|=1$. This leads to a transfer window of width $\sim 1$ around $k=0$. The scattering phase shift at zero energy, $\zeta_1=\ii\frac{1}{t_1}\tfrac{dt_1}{dk}\big|_{k=0}$, evaluates numerically to $\zeta_1=2.014$. In the opposite limit of $|k|\gg 1$, the two channels effectively decouple.

\begin{figure*}[t]
\includegraphics[width=\textwidth]{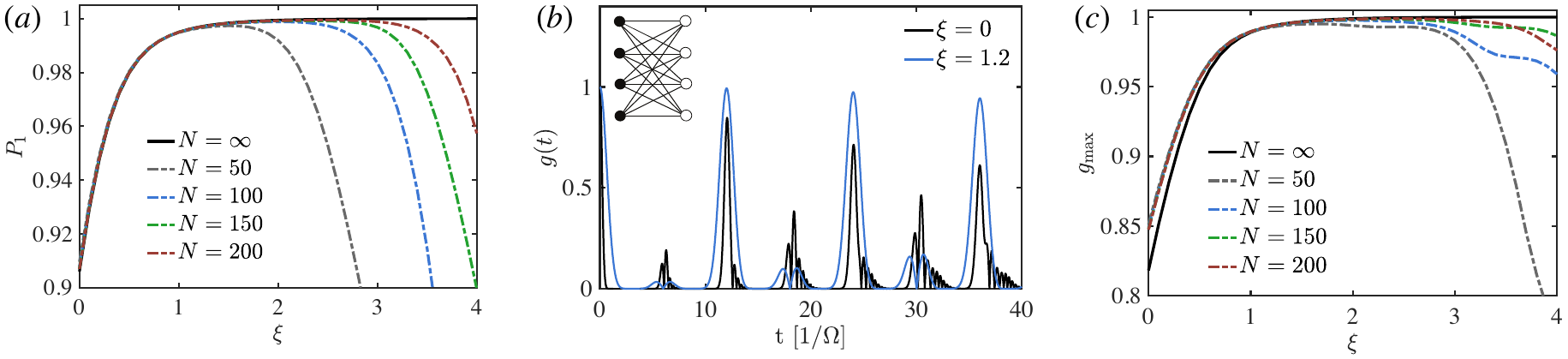}
  \caption{(a) Maximal revival fidelity $g_\mathrm{max}$ and (c) transmission probability $P_1$ as a function of the squeezing parameter $\xi$ for various system sizes. We note a finite-size effect at high squeezing, which occurs when squeezing expands the spin coherent state around a great circle of a finite-radius Bloch sphere. (b) Many-body fidelity $g(t)$ for a CBG with $N=200$, initialised in both an unsqueezed ($\xi=0$) and a squeezed ($\xi=1.2$) initial $\mathbb{Z}_2$ state.}
    \label{fig:complete_squeezing}
\end{figure*}

The initial state of interest here, $\ket{\mathbb{Z}_2^A}$, corresponds to an incoming Gaussian wavepacket in channel $A$, i.e., $\braket{x}{A(t=0)}=\pi^{-1/4}\ee^{-(x-S\pi)^2/2}$ and $\braket{x}{B(t=0)}=0$. Its decomposition into scattering eigenstates is thus also Gaussian, with Fourier amplitudes $\tilde{A}(k)=2^{1/2}\pi^{1/4}\ee^{-k^2/2}$. The probability
\begin{align}
    P_1=\frac{1}{2\pi}\int_{-\infty}^\infty|t_1(k)\tilde{A}(k)|^2dk
\end{align}
of a change of channel at the first scattering event evaluates to $P_1\approx0.906$. After the scattering event, one can again apply the decoupling approximation, to propagate the wave-function between subsequent scattering events at $t\approx (2\mathbb{N}+1)\pi$, showing that a large value for $P_1$ is required for the expected alternating (approximate) revivals of the two $\mathbb{Z}_2$ states. The quality of these revivals is quantified by the many-body fidelity $     g(t)=|\braket{\psi(0)}{\psi(t)}|\,$, which is displayed in Fig.~\ref{fig:complete_squeezing}. It shows alternating smaller and larger peaks, where the former (at $\Omega t\approx 2\pi$) result from the finite probability of the system not changing channel in the scattering process ($1-P_1\approx0.094$), while the latter (at $\Omega t\approx 4\pi$) appears due to two consecutive scattering processes resulting in two successive changes of channel. The height of this first (large) revival of $g(t)$, which we denote $g_\mathrm{max}$, can thus be calculated as
\begin{align}
    g_\mathrm{max}=\bigg|\frac{1}{2\pi}\int_{-\infty}^\infty\ee^{\ii\tau}\tilde{A}_2(k)\tilde{A}(k)^*dk\bigg|\,,
\end{align}
where $\tilde{A}_2(k)=(t_0(k)^2+t_1(k)^2)\tilde{A}(k)$ is the wavefunction of the state on bipartition $A$ after two scattering events and the phase $\tau$ is introduced to account for the phase shift acquired by $\tilde{A}_2(k)$ relative to $\tilde{A}(k)$. Since $|t_1(k)^2\tilde{A}(k)|\gg|t_0(k)^2\tilde{A}(k)|$, as a first approximation we may choose $\tau=2\zeta_1$ for which $g_\mathrm{max}\approx0.818$. The exact phase shift maximizing $g_\mathrm{max}$ can be determined numerically as $\tau=3.55$ and results in $g_\mathrm{max}\approx0.846$. The fact that $g_\mathrm{max}<1$ reflects the fact that the PXP models host only approximate QMBS.

\textit{Squeezed QMBS.---}
The scattering picture directly suggests the possibility of constructing different initial states which will display enhanced revivals. Specifically, the transfer window at $k=0$ (see Fig.~\ref{fig:scattering_problem}) indicates that $P_{1}$ and $g_{\rm max}$ could be increased by squeezing the width of the initial state momentum distribution $\tilde A(k)$ below $1$. A natural candidate is the class of spin squeezed states~\cite{ma_quantum_2011,gross_spin_2012,kitagawa_squeezed_1993}, defined as $\ket{S,\theta,\varphi;\chi}=V_\chi\ket{S,\theta,\varphi}$ with
\begin{align}
    \label{eq:spin squeezed state}
    V_\chi=\ee^{\chi/2\left((\hat{S}^+)^2-(\hat{S}^-)^2\right)}\,.
\end{align}
A squeezed $\mathbb{Z}_2$ initial state $\ket{\psi(0)}=V_\chi\otimes \mathbb{1}\ket{\mathbb{Z}_2^A}$ thus corresponds to a 
squeezed Fourier wavepacket  $\tilde{A}(k)=2^{1/2}\pi^{1/4}\ee^{-k^2\ee^{2\xi}/2+\xi/2}$, with $\xi=\chi N/2$. As shown in Fig.~\ref{fig:complete_squeezing}, squeezing the  initial states indeed  leads to larger and more long-lived revivals with increasing $\xi$, (until finite system size effects become relevant at $\xi\sim \log(N)$).

Even though the scattering problem is rigorously defined only on the infinite CBG,  we find  that  spin squeezing of the initial $\mathbb{Z}_2$ states leads to enhanced revivals in PXP models also on other bipartite graphs. This requires a generalization of the spin squeezing operator. A physically motivated choice that takes into account the locality on a given bipartite graph is the quasi-local transformation
\begin{align}
\label{eq:local squeezer}
    U_\chi=\ee^{\chi/2\sum_{i\in B}\left((\hat{S}^+_i)^2-(\hat{S}_i^-)^2\right)}\,.
\end{align}
Here, the $\hat{S}_i^\pm$ denote the ladder operators for the compound spin systems of the nearest neighbours of the $i$-th site. For instance, in the case of a circle of $N$ sites,
\begin{align}
    \hat{S}_i^\pm=\frac{1}{2}\sigma_{i-1}^\pm+\frac{1}{2}\sigma_{i+1}^\pm\,,
\end{align}
where the addition in the subscript is modulo $N$. We will refer to this as local squeezing, while we dub the transformation~\eqref{eq:spin squeezed state} global squeezing. The physical reasoning underlying this choice of squeezing operator is the following. Each spin on partition $B$ is initially blockaded by its $z$ nearest neighbours and frozen in the state $\ket{0}_i$. From the perspective of this spin, this is analogous to the situation in a CBG of size $2z$. Based on our results for the CBG, we may expect that squeezing of the initial state of its nearest neighbours liberates that particular spin most efficiently, resulting in more pronounced revivals. Aplying this argumentto all spins in partition $B$ implies~\eqref{eq:local squeezer}.

\begin{figure}[t]
\hspace*{-1.6cm}
\includegraphics[width=0.37\textwidth]{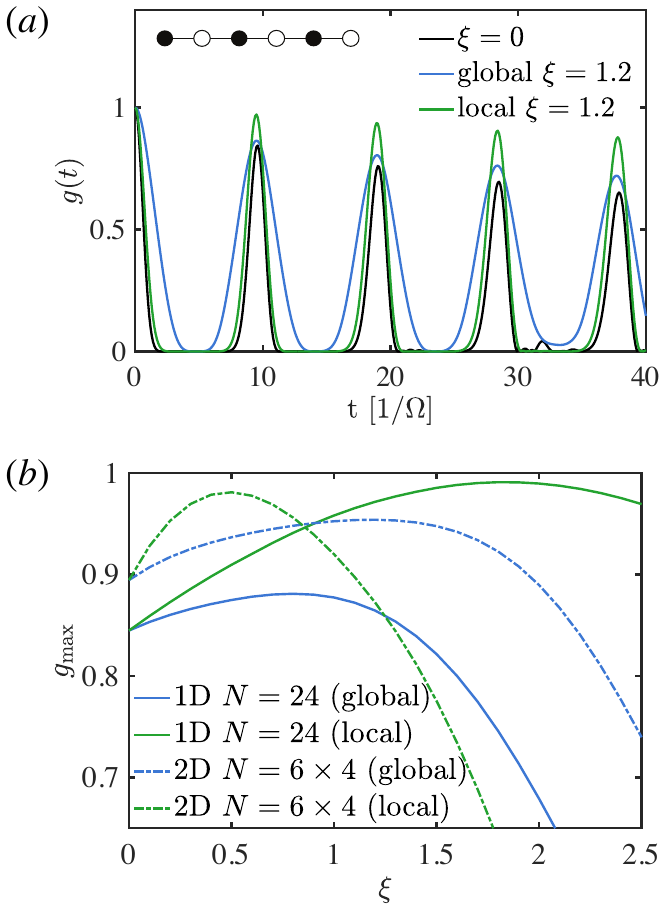}
  \caption{(a) Many-body fidelity $g(t)$ for a 1D circle with $N=24$, initialised in both an unsqueezed ($\xi=0$) and globally and locally squeezed ($\xi=1.2$) initial $\mathbb{Z}_2$ state. (b) First maximum of the many-body fidelity $g_\mathrm{max}$ as a function of squeezing parameter $\xi$, for initial $\mathbb{Z}_2$ states squeezed both globally and locally, on 1D and 2D lattices with $N=24$ sites.}
    \label{fig:1D_and_2D_squeezing}
\end{figure}

We solve for the dynamics of squeezed initial states in the 1D chain and 2D square lattice (with periodic boundary conditions) by numerical integration of the corresponding Schr\"odinger equation on finite system sizes, and compare the local squeezing approach \eqref{eq:local squeezer} with a naive application of a global squeezing operation (see Fig.~\ref{fig:1D_and_2D_squeezing}). For small squeezing, both approaches result in enhanced and more long-lived revivals with increasing $\xi$. Importantly, the local squeezing gives significantly better results, leading to almost perfect revivals and thus supporting the physical reasoning outlined above. 

\textit{Hamiltonian deformations.---}
Our interpretation of the PXP dynamics in terms of a scattering problem has motivated deformations of the initial states in order to enhance the periodic dynamics associated with QMBS in the PXP model. In a complementary view, one might instead deform the Hamiltonian and leave the initial state invariant. Specifically, instead of squeezing the initial state, we can in a unitarily equivalent way conjugate the Hamiltonian with the squeezing operator to obtain $H_{\chi,\mathrm{global}}=V_\chi^\dag H V_\chi$ or $H_{\chi,\mathrm{local}}=U_\chi^\dag H U_\chi$, respectively. Clearly, these transformed Hamiltonians will induce enhanced revivals with increasing $\chi$, that now occur for the plain $\ket{\mathbb{Z}_2^A}$ initial state rather than its squeezed counterpart. Thus our analysis ultimately allows us to construct a deformation of the PXP Hamiltonian that gives rise to perfect revivals on the infinite CBG and significantly enhanced revivals on other lattices.
For instance, in 1D the deformed Hamiltonian is
\begin{align}
\label{eq:deformation}
    H_{\chi,\mathrm{local}}=H+\frac{\Omega\chi}{2}\sum_i\mathcal{P}\sigma^x_i\mathcal{P}\sigma^z_{i+2}+O(\chi^2)\,.
\end{align}
Remarkably, the leading order terms introduced by this deformation are of the same form as those studied in Refs.~\cite{khemaniSignaturesIntegrabilityDynamics2019,choiEmergentSUDynamics2019}, where it was shown that they lead to virtually perfect QMBS. However, the physical motivation for the particular form of such deformations has remained uncertain. In contrast, these deformations emerge naturally in the scattering picture developed in this work. This highlights one of the main merits of the scattering approach. 

\textit{Discussion.---}
In this work we have discussed the phenomenon of QBMS in PXP models, developing a new analytical approach to studying QMBS based on an equivalence between the PXP model on the CBG and a 1D chiral scattering problem. The resulting novel insights into the mechanism underlying scars in PXP models allow the conclusion that dynamical signatures of QMBS can be enhanced by spin squeezing. This could prove a valuable tool for quantum information processing tasks based on quantum many-body scars \cite{dooleyRobustQuantumSensing2021}, and could potentially be combined with other approaches of actively stabilising QMBS, for instance through driving~\cite{bluvsteinControllingQuantumManybody2021,michailidisStabilizingTwodimensionalQuantum2020,mukherjeeCollapseRevivalQuantum2020,sugiuraManybodyScarState2021}. 

While our focus on PXP models on CBGs is motivated by the analytical tractability, we note that such models could be potentially realized in Rydberg atom array experiments. For instance, in a two-species setting~\cite{singh_dual-element_2021} with appropriately chosen Rydberg states, weak intraspecies interactions and strong interspecies interactions could result in an effective PXP model on the CBG where the two atomic species represent the two partitions.  

\begin{acknowledgments}
We thank M.~Endres, S.~Choi, W.~W.~Ho, G.~Giudici and H.~Bernien for useful discussions. This work is supported by the Erwin Schr\"odinger Center for Quantum Science and Technology  through a Discovery Grant. 
\end{acknowledgments}

\appendix
\section{Spin Coherent States}
\label{app:coherent states}

\begin{figure*}
    \centering
    \includegraphics[width=.85\textwidth]{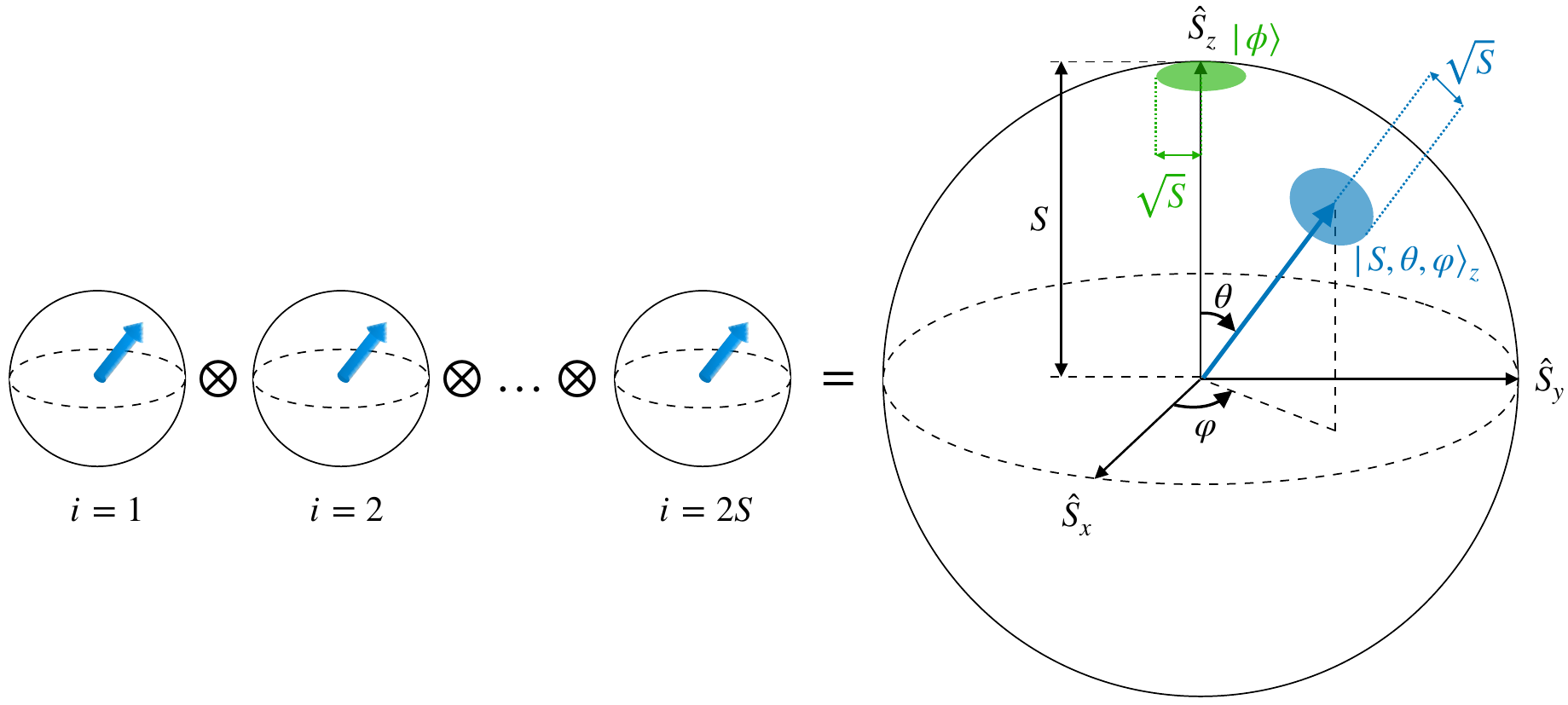}
    \caption{Generalised Bloch sphere for an arbitrary spin coherent state $\ket{S,\theta,\varphi}$ for an ensemble of $2S$ spin-$1/2$ systems (represented by unit Bloch spheres).}
    \label{fig:generalised_Bloch}
\end{figure*}

\subsection{Bosonic coherent states}\label{app:bosonic coherent states}
Before discussing spin coherent states, we recall (very briefly) the class of \textsl{bosonic coherent states}, states with a well-defined phase rather than photon number~\cite{walls_quantum_2008,gerry_introductory_2005,gazeau_coherent_2009}. We consider a Hilbert space constructed as a Fock space upon some vacuum $\ket{0}$ using the bosonic creation and annihilation operators, $\hat{a}^\dagger$ and $\hat{a}$, which satisfy $[\hat{a},\hat{a}^\dagger]=1$, $[\hat{a},\hat{a}]=[\hat{a}^\dagger,\hat{a}^\dagger]=0$ and $\hat{a}\ket{0}=0$. The Hilbert space is spanned by the orthonormal basis of number states $\ket{n}$ ($n=0,1,2,\dots$). The phase space associated to this system is spanned by quadrature operators
\begin{align}
\begin{split}
    \label{eq:quadratures}
    \hat{x}&=\frac{\ii}{\sqrt{2}}(\hat{a}^\dagger-\hat{a})=\sqrt{2}\mathrm{Im}(\hat{a})\\
    \hat{p}&=\frac{1}{\sqrt{2}}(\hat{a}^\dagger+\hat{a})=\sqrt{2}\mathrm{Re}(\hat{a})\,,
\end{split}
\end{align}
which satisfy $[\hat{x},\hat{p}]=-\ii$. On this Hilbert space, coherent states $\ket{\alpha}$ ($\alpha\in\mathbb{C}$) are generated as
\begin{align*}
    \ket{\alpha}=\mathcal{D}(\alpha)\ket{0}\,,\quad\mathcal{D}(\alpha)=\exp\left(\alpha\hat{a}^\dagger-\alpha^*\hat{a}\right)\,.
\end{align*}
Here, $\mathcal{D}$ displaces a state in the phase space, without changing its shape (i.e. the shape of its wavefunction, phase space distribution, etc.), which means that coherent states are minimum uncertainty states ($\Delta x\Delta p=1/2$) and more specifically $\Delta x=\Delta p$. This can be seen directly from the fact that
\begin{align*}
    &\mathcal{D}(\alpha)^{-1}\hat{a}\mathcal{D}(\alpha)=\hat{a}+\alpha\\
    &\mathcal{D}^{-1}(\alpha)\hat{a}^\dagger\mathcal{D}(\alpha)=\hat{a}^\dagger+\alpha^*\,,
\end{align*}
i.e. we can write the displacement parameter $\alpha$ as $\alpha=\frac{1}{\sqrt{2}}\left(p_0+\ii x_0\right)$ where $(x_0,p_0)$ are the coordinates of the coherent state in phase space. Viewing the phase space as an Argand diagram with real axis $\hat{x}$ and imaginary axis $\hat{p}$, the displacement vector for state $\ket{\alpha}$ is $\ii\alpha^*$ in our convention. Of course, the vacuum itself is a coherent state with $\alpha=0$.

We can calculate the overlap of coherent states $\ket{\alpha}$ and $\ket{\beta}$ as
\begin{align}
    \label{eq:bosonic overlap}
    \braket{\beta}{\alpha}&=\exp\left(-\frac{1}{2}\left(|\alpha|^2+|\beta|^2\right)+\alpha\beta^*\right)\,.
\end{align}
Note that if we insert the expression for $\alpha$, the form of these overlaps is Gaussian in $x$ and $p$, with a phase that is dependent on the phase space coordinates. Also note that we can expand a coherent state in the basis of number states to arrive at the alternative definition
\begin{align}
\label{eq:alternative bosonic coherent}
    \ket{\alpha} &=\exp\left(-\frac{1}{2}|\alpha|^2\right)\exp\left(\alpha\hat{a}^\dagger\right)\ket{0}\,.
\end{align}

\subsection{Spin coherent states}\label{app:spin coherent states}

\subsubsection{Background: group-theoretic coherent states}
In mathematical physics, the traditional bosonic coherent states are subsumed into a much larger class of group theoretical coherent states. The basic ingredients for a group-theoretic coherent state are a reference (vacuum) state and a displacement operator, just as for the bosonic case. In the group-theoretic case, the set of displacement operators form a projective unitary representation of a semi-simple Lie group~\cite{guaita_generalization_2021}.

Apart from the bosonic coherent states, the group-theoretic coherent states associated to the group $\mathrm{SU}(2)$ are most well-known in physics as the \textsl{spin coherent states}. On some Hilbert space $\mathcal{H}$, they are defined with respect to a vacuum state via a unitary representation of $\mathrm{SU}(2)$ on $\mathcal{H}$. Since $2S+1$-dimensional unitary representations of $\mathrm{SU}(2)$ on some $\mathcal{H}$ are associated with a spin-$S$ system, spin coherent states represent a certain subset of the states in a spin-$S$ Hilbert space.

\subsubsection{Spin coherent states of a spin-S system}
For a spin-$S$ system, the $2S+1$-dimensional unitary representation of $\mathrm{SU}(2)$ on the spin system Hilbert space is realised by the operators $\hat{S}^\mu$ ($\mu=x,y,z$) and $\hat{S}^2=\sum_\mu(\hat{S}^\mu)^2$, which satsify the $SU(2)$ operator algebra $[\hat{S}^\mu,\hat{S}^\nu]=\ii\sum_\omega\epsilon^{\mu\nu\omega}\hat{S}^\omega$. We can express the state of the system in the basis of $S_\mu$-eigenstates which we label $\ket{S,m}_\mu$ ($m\in\mathbb{Z}$, $-S\leq m\leq +S$) and which satisfy
\begin{align*}
    &\hat{S}^\mu\ket{S,m}_\mu=m\ket{S,m}_\mu\\
    &\hat{S}^2\ket{S,m}_\mu=S(S+1)\ket{S,m}_\mu\,,
\end{align*}
and ${}_\mu\langle S,m'|S,m\rangle_\mu=\delta_{mm'}$. The ground state of the spin system is $\ket{\phi}\equiv\ket{S,-S}_z$. With respect to this state, we now define the class of spin coherent states $\ket{\eta}$ ($\eta\in\mathbb{C}$) according to~\cite{radcliffe_properties_1971}
\begin{align*}
    \ket{\eta}=\frac{1}{(1+|\eta|^2)^S}\exp\left(\eta\hat{S}^+\right)\ket{\phi}\,.
\end{align*}
Comparing this
to~\eqref{eq:alternative bosonic coherent}, the analogy to the bosonic states is immediately obvious: here, $\ket{\phi}$ takes the role of the reference vacuum state and $\eta$ parametrises the spin coherent states in the same way as $\alpha$. We can show that in the spin-$z$ eigenstate basis
\begin{align}
    \label{eq:spin coherent defn 1}
    \ket{\eta}=\frac{1}{(1+|\eta|^2)^S}\sum_{m=-S}^{+S}{2S\choose S+m}^{1/2}\eta^{S+m}\ket{S,m}_z\,.
\end{align}
The state takes a similar form in the other spin component eigenbases.

\subsubsection{Spin coherent states of an atomic ensemble}
One way in which a spin-$S$ Hilbert space may arise is as the Hilbert space of $N=2S$ spin-$1/2$ systems with internal states $\{\ket{0},\ket{1}\}$. For $N$ spin-$1/2$ systems, labelled by $i=1,\dots,N$ and with individual spin operators
\begin{align*}
    \hat{s}_\mu=\frac{1}{2}\sigma_\mu^{(i)}~(\mu=x,y,z)\,,\quad\hat{s}^2=\hat{s}_x^2+\hat{s}_y^2+\hat{s}_z^2\,,
\end{align*}
we define the Spin operators $\hat{S}^\mu$ ($\mu=x,y,z$) and $\hat{S}^2$ specifically as the compound spin operators
\begin{align*}
    \hat{S}^\mu=\frac{1}{2}\sum_{i=1}\sigma^\mu_i\,,\quad \hat{S}^2=(\hat{S}^x)^2+(\hat{S}^y)^2+(\hat{S}^z)^2\,.
\end{align*}
In terms of the underlying atomic lattice, in this context the $\ket{S,m}_\mu$ are so-called \textsl{Dicke states} and are defined as equal superpositions of all individual configurations of the atoms that lead to a specific value of $m$~\cite{garraway_dicke_2011}. The collective ground state of the array is still $\ket{\phi}=\ket{S,-S}_z=\bigotimes_{i=}^N\ket{0}_i$. The formal definition of the spin coherent states of an atomic ensemble follows naturally. We can also compute the spin coherent state overlaps, in analogy to~\eqref{eq:bosonic overlap}, as~\cite{radcliffe_properties_1971}
\begin{align}
    \label{eq:spin overlap 1}
    \langle\lambda|\eta\rangle &=\frac{(1+\lambda^*\eta)^{2S}}{(1+|\lambda|^2)^S(1+|\eta|^2)^S}\,.
\end{align}

Rather than in the basis of Dicke states, one could express the state of the atomic ensemble in the form $\ket{\Theta,\Phi}=\otimes_{i=1}^N\ket{\theta_i,\varphi_i}_i$ with spin-$1/2$ states
\begin{align*}
    \ket{\theta_i,\varphi_i}_i\equiv\cos(\theta_i/2)\ket{0}_i+\ee^{-\ii\varphi_i}\sin(\theta_i/2)\ket{1}_i\,,
\end{align*}
where $\Theta\equiv(\theta_1,\dots,\theta_N)$ and $\Phi\equiv(\varphi_1,\dots,\varphi_N)$ and where $(\theta_i,\varphi_i)$ parametrise the Bloch sphere of the $i$-th spin-$1/2$ system in the familiar way. Clearly, $\ket{\phi}$ then corresponds with the state where $\theta_i=\varphi_i=0$. In close analogy to the bosonic coherent states, we then define the class of spin coherent states as states of fixed phase, i.e. those states with $\theta_i=\theta$ and $\varphi_i=\varphi$ for all $i$,
\begin{align*}
    \ket{S,\theta,\varphi}\equiv\bigg(\cos(\tfrac{\theta}{2})\ket{0}+\ee^{-\ii\varphi}\sin(\tfrac{\theta}{2})\ket{1}\bigg)^{\otimes N/2}\,.
\end{align*}
In a Dicke basis, these states take the form
\begin{widetext}
\begin{align}
    \label{eq:spin coherent defn 2}
    \ket{S,\theta,\varphi}\equiv\bigotimes_{i=1}^N\ket{\theta,\varphi}_i=\sum_{m=-S}^{+S}{2S\choose S+m}^{1/2}\cos^{2S}(\theta/2)\tan^{S+m}(\theta/2)\ee^{-\ii\varphi(S+m)}\ket{S,m}_z\,.
\end{align}
\end{widetext}
Cleary, the definitions in~\eqref{eq:spin coherent defn 1} and~\eqref{eq:spin coherent defn 2} are identical for $\eta=\tan(\theta/2)\ee^{-\ii\varphi}$. This also allows us to reproduce the expressions for the overlaps,~\eqref{eq:bosonic overlap} and~\eqref{eq:spin overlap 1}, in this new parametrisation, and for instance $\langle\phi|S,\theta,\varphi\rangle=\cos^{2S}(\theta/2)$~\cite{radcliffe_properties_1971}. The parametrisation in terms of the angles $(\theta,\varphi)$ allows us to represent the spin coherent states $\ket{S,\theta,\varphi}$ as lying on a \textsl{generalised Bloch sphere} of radius $S$: it can be shown that the state $\ket{S,\theta,\varphi}$ is equivalently given by $\ket{S,\theta,\varphi}=R(\theta,\varphi)\ket{\phi}$, where $R(\theta,\varphi)$ is the rotation operator on the generalised Bloch sphere~\cite{ma_quantum_2011} (see Fig.~\ref{fig:generalised_Bloch}).

\section{Holstein-Primakoff Transformation}\label{app:Holstein-Primakoff Transformation}

\begin{figure*}
    \centering
    \includegraphics[width=.85\textwidth]{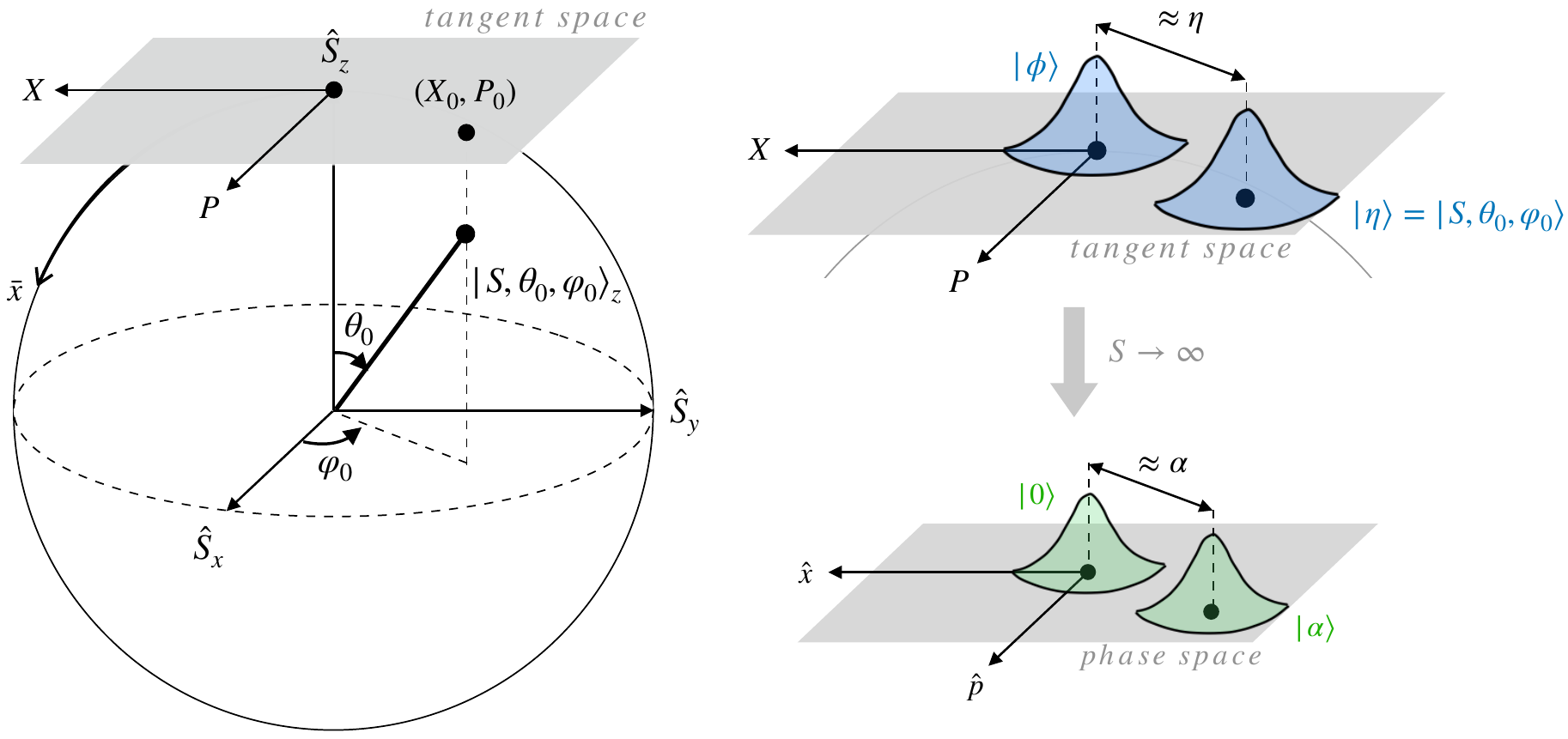}
    \caption{Geometry of the Holstein-Primakoff transformation on the generalised Bloch sphere. The left figure shows the parametrisation of the tangent space to the generalised Bloch sphere in Fig.~\ref{fig:generalised_Bloch} at $\theta=\varphi=0$ in terms of coordinates $(X,P)$, as well as some spin coherent state $\ket{S,\theta_0,\varphi_0}$. The right hand side shows the relationship between the tangent space and bosonic phase space as described in the main text. Note that the shading of the coherent states is purely illustrative of the shape of spin coherent states on phase space and should not be taken too literally (since it would contradict $\braket{\phi}{\eta}\sim\mathrm{O}(1)$).}
    \label{fig:Holstein_Bloch}
\end{figure*}

\subsection{Operator transformation}\label{app:Holstein operators}
We now want to consider the way in which we can map the spin basis for the coherent states to a Fock basis that spans the subspace of a bosonic Hilbert space, which grows in size relative to the embedding Hilbert space as $S$ increases. We follow here closely the reasoning and the notation in~\cite{klein_boson_1991}.

We begin by mapping the states $\{\ket{S,m}_z\}_{-S\leq m\leq S}$ onto the states $\{\ket{S,n-S}_z\}_{0\leq n\leq2S}$ and using the shorthand notation $\ket{S,n-S}_z\equiv\ket{n}$. We can then write
\begin{align}
&\langle n|\hat{S}^z|n \rangle=n-S \\
&\langle n+1|\hat{S}^+|n\rangle=\sqrt{(2S-n)(n+1)}\,.
\end{align}
Comparing this with the matrix elements for the bosonic creation and annihilation operators $\hat{a},\hat{a}^\dagger$ satisfying the commutation relation $[\hat{a},\hat{a}^\dagger]=1$ when treating the states $\ket{n}$ as a Fock space basis,
\begin{align}
&\langle n|\hat{a}^\dagger\hat{a}|n \rangle=n \\
&\langle n+1|\hat{a}^\dagger|n\rangle=\sqrt{n+1}\,,
\end{align}
we are naturally lead to the operator transformation $\hat{S}^z=\hat{a}^\dagger\hat{a}-S$ and $\hat{S}^+=\hat{a}^\dagger\sqrt{2S-\hat{a}^\dagger\hat{a}}$. In the limit of $S\to\infty$, the second transformation becomes $\hat{S}^+\to\hat{a}^\dagger\sqrt{2S}$. From an operator algebraic perspective, the interpretation of this is straightforward: We can treat the states $\ket{n}$ as a basis of a subspace in the Fock space constructed on the vacuum $\ket{n=0}=\ket{S,-S}_z$, up to excitations $n=2S$. Of course, in the limit $n\to\infty$, this basis becomes infinite and so the complement of the subspace in the embedding bosonic Hilbert space becomes negligible; we can see that in fact the spin ladder operators in this limit simply become the (re-scaled) bosonic creation and annihilation operators. The Holstein-Primakoff transformation, in summary, states that
\begin{align}
    \label{eq:operator transformation}
    \lim_{S\to\infty}\hat{S}^+=\hat{a}^\dagger\sqrt{2S}\,,\quad\lim_{S\to\infty}\hat{S}^-=\hat{a}\sqrt{2S}
\end{align}
which also implies (see~\eqref{eq:quadratures}) that
\begin{align}
    \lim_{S\to\infty}\hat{S}^x=\hat{p}\sqrt{S}\,,\quad
    &\lim_{S\to\infty}\hat{S}^y=-\hat{x}\sqrt{S}\,.
\end{align}

\subsection{Transformation of spin coherent states}\label{app:Holstein geometry}
One particularly interesting aspect of the Holstein-Primakoff Transformation is the way in which it transforms the spin coherent states of a spin-$S$ system in the limit of $S\to\infty$. We assume ad hoc for now that in this limit, $\eta\to\alpha/\sqrt{2S}$, which we will motivate momentarily when considering the geometric interpretation of the transformation. We then have that
\begin{align}
   & \lim_{S\to\infty}\left(1+|\eta|^2\right)^{-S}=\exp\left(-\frac{1}{2}|\alpha|^2\right)\,, \\
    &\lim_{S\to\infty}\exp\left(\eta\hat{S}^+\right)\ket{\phi}=\exp\left(\alpha\hat{a}^\dagger\right)\ket{0}\,.
\end{align}
This tells us that
\begin{align}
    \lim_{S\to\infty}\ket{\eta}=\exp\left(-\frac{1}{2}|\alpha|^2\right)\exp\left(\alpha\hat{a}^\dagger\right)\ket{0}\,,
\end{align}
which is identically the definition of the bosonic coherent state according to~\eqref{eq:alternative bosonic coherent}. This means that the Holstein-Primakoff transformation transforms spin coherent states to bosonic coherent states.

There is also a geometric interpretation of this transformation. Consider the generalised Bloch sphere of finite radius $S$ and consider the tangent space to the Bloch sphere at $\theta=\varphi=0$ spanned by coordinates $(X,P)$ as shown in Fig.~\ref{fig:Holstein_Bloch}. In this tangent space, the ground state $\ket{\phi}$ has coordinates $X=P=0$. Now consider those spin coherent states $\ket{S,\theta,\varphi}$ which satisfy $\theta\ll1$, i.e. states for which $\braket{\phi}{S,\theta,\varphi}\sim O(1)$, meaning these states lie close to the north pole of the Bloch sphere. Such states have tangent space coordinates
\begin{align}
    X&=-S\sin(\theta)\sin(\varphi)\\
    P&=S\sin(\theta)\cos(\varphi)\,.
\end{align}
However now note that Fig.~\ref{fig:Holstein_Bloch} shows that $X=-\hat{S}^y$ and $P=\hat{S}^x$. This is a somewhat heuristic comparison with slight abuse of notation, since we are not clearly distinguishing between geometric axes and operators, however the notion should be self-explanatory and the equations could, for instance, be interpreted in terms of the eigenvalues of the spin operators. Eq.~\eqref{eq:operator transformation} therefore implies that in the limit $S\to\infty$, the tangent space coordinates $(X,P)\to(\hat{x}\sqrt{S},\hat{p}\sqrt{S})$. This means that in this limit, the tangent space to the Bloch sphere of the spin system becomes the (re-scaled) phase space of the bosonic system to which it is mapped.

Note also the arc length from the north pole in the direction of decreasing $\theta$, $\bar{x}=-S\theta$. From Fig.~\ref{fig:Holstein_Bloch}, it is evident that $\bar{x}\approx X$ for $\bar{x}\ll1$ and that $\bar{x}\to X$ exactly in the limit $S\to\infty$.

For the purpose of this paper, the most sensible way to look at this geometric picture is to consider a specific spin coherent state $\ket{S,\theta_0\ll1,\varphi_0}$ as the radius of curvature of the Bloch sphere tends to infinity (i.e. the sphere flattens out) in the limit $S\to\infty$. Clearly, in this limit the state will lie on the tangent space to the Bloch sphere at coordinates $(X_0,P_0)$ (see Fig.~\ref{fig:Holstein_Bloch}). Now consider the parameter $\eta(\theta_0,\varphi_0)=\tan(\theta_0)\exp(-\ii\varphi_0)$ for this coherent state. Evidently,
\begin{align}
\begin{split}
    \eta(\theta_0,\varphi_0)
    &\approx
    \frac{1}{2S}\left(P_0+\ii X_0\right)\,.
\end{split}
\end{align}
This means that
\begin{align}
    \lim_{S\to\infty}\eta=\frac{1}{2\sqrt{S}}(p_0+\ii x_0)=\frac{\alpha}{\sqrt{2S}}
\end{align}
where $\alpha$ and $(x_0,p_0)$ equivalently describe the position of the bosonic coherent state $\ket{\alpha}=\lim_{S\to\infty}\ket{\eta}$ as previously discussed. This is exactly the assumption that we made at the top of this section and we see now that it is validated by the geometric picture.

\subsection{Spin coherent state wavefunctions}\label{app:Holstein wavefunctions}

\subsubsection{Position-space basis}
To carry out the calculations in the main text, we will need an explicit expression for the position-space wavefunction of the vacuum state $\ket{\phi}$ (and by extension for states $\ket{S,\theta,\pi/2}$). To do this, we refer to the previous analysis of the Holstein-Primakoff Transformation to note that in the limit $S\to\infty$, we can identify the eigenstates of the operators $\hat{x}$ and $\hat{S}^y$ according to $\ket{S,m}_y\to\ket{x=-m/\sqrt{S}}$. We also note that~\eqref{eq:spin coherent defn 2} implies that
\begin{align}
    \ket{\phi}=\frac{1}{2^S}\sum_{m=-S}^{+S}{2S\choose S+m}^{1/2}\ket{S,m}_y\,.
\end{align}
Taking the limit of this expression as $S\to\infty$, we make use of the fact that the limit of a binomial distribution is Gaussian, in the sense that
\begin{align}
\lim_{n\to\infty}p^s(1-p)^{n-s}{n\choose s}=\frac{\exp\left(-\frac{(s-np)^2}{2np(1-p)}\right)}{\sqrt{2\pi p(1-p)n}}\,.
\end{align}
We now consider the quantity $|\langle S,m|\phi\rangle|^2$, which will clearly indeed be binomial. Therefore, setting $p=1/2$, $n=2S$ and $s=S+m$ in the above limit,
\begin{align}
    \label{eq:phi wavefunction}
    \lim_{S\to\infty}\ket{\phi}=\int_{-\infty}^\infty dx~\phi(x)\ket{x}
\end{align}
with continuous coefficients
\begin{align}
    \phi(x)=\frac{1}{\pi^{1/4}}\ee^{-x^2/2+\ii\beta(x)}\,,\quad\beta\in\mathbb{R}\,.
\end{align}
Clearly, this defines a wavefunction for $\phi(x)$. In fact, we will choose the convention that the wavefunction of this and all other spin coherent states will be real ($\beta=0$). While this is not strictly speaking accurate, it will suffice in this case, since (as we will see) we will only really be interested in the Fourier transform of the spin coherent state wavefunctions (i.e. the momentum-space wavefunctions), on which a phase will have no effect. Therefore we write the position-space wavefunction for a general spin coherent state as 
\begin{align}
\label{eq:spin coherent wavefunctions - x}
\braket{x}{S,x_0}&=\frac{1}{\pi^{1/4}}\ee^{-(x-x_0)^2/2}\,,
\end{align}
where we use the shorthand notation $\ket{S,x_0}$ to denote a spin coherent state at $x=x_0$ along the great circle with $\varphi=\pi/2$.

\subsubsection{Momentum-space basis}

We define the Fourier transform between conjugate variables $x,k$ for the purposes of this paper as
\begin{align}
    A(x) &= \frac{1}{2\pi}\int_{-\infty}^\infty \tilde{A}(k)\ee^{-\ii k x}dk \\
    \tilde{A}(k) &= \int_{-\infty}^\infty A(x)\ee^{\ii k x}dx\,.
\end{align}
Hence, the momentum-space wavefunction associated with~\eqref{eq:spin coherent wavefunctions - x} reads
\begin{align}
\label{eq:spin coherent wavefunctions - k}
\braket{k}{S,x_0}&=2^{1/2}\pi^{1/4}\ee^{-k^2/2}\,.
\end{align}

\section{Scattering coefficients calculation}
\label{app:calculation}

In this appendix, we discuss in detail the solution to the coupled differential equations, leading ultimately to the expressions for the scattering coefficients. We also provide some analytical expressions for the (iterated) Gaussian integrals $G(k)$, $H(k,x)$ (and $H(k)$) and $F(k)$.

\subsection{Formal solution}
We start from the differential equations derived in the main text,
\begin{equation}
\begin{aligned}
    A'(x)&=\ii kA(x)-\langle\phi|B\rangle~\phi'(x)\\
    B'(x)&=\ii kB(x)-\langle\phi|A\rangle~\phi'(x)\,.
\end{aligned}    
\end{equation}
We can solve these equations formally by integrating, to obtain
\begin{align*}
    A(x)&=A(x_0)\ee^{\ii k(x-x_0)}-\langle\phi|B\rangle \int_{x_0}^x\phi'(z)\ee^{-\ii kz}~dz\\
     B(x)&=B(x_0)\ee^{\ii k(x-x_0)}-\langle\phi|A\rangle \int_{x_0}^x\phi'(z)\ee^{-\ii kz}~dz\,.
\end{align*}
We now consider scattering boundary conditions corresponding with a plane wave incident from $x_0\ll0$ on the active partition $A$, such that $A(x)=A(x_0)\ee^{\ii k(x-x_0)}$ and $B(x_0)=0$, whereby
\begin{align*}
    A(x)&=A(x_0)\ee^{\ii k(x-x_0)}-\langle\phi|B\rangle \int_{x_0}^x\phi'(z)\ee^{-\ii kz}~dz\\
    B(x)&=-\langle\phi|A\rangle \int_{x_0}^x\phi'(z)\ee^{-\ii kz}~dz\,.
\end{align*}
We can use this general solution to calculate self-consistent expressions for the coefficients $\langle\phi|A\rangle$ and $\langle\phi|B\rangle$ by taking the overlap with $\ket{\phi}$, so that
\begin{align*}
\langle\phi|A\rangle&=A(x_0)\int_{-\infty}^\infty\phi(x)\ee^{\ii k(x-x_0)}dx\\
&-\langle\phi|B\rangle\int_{-\infty}^\infty\phi(x)\int_{x_0}^x\ee^{\ii k(x-z)}\phi'(z)~dz~dx\\
\langle\phi|B\rangle &=-\langle\psi|A\rangle\int_{-\infty}^\infty\phi(x)\int_{x_0}^x\ee^{\ii k(x-z)}\phi'(z)~dz~dx\,.
\end{align*}
We can substitute each of these equations into the other to obtain independent expressions for $\langle\phi|A\rangle$ and $\langle\phi|B\rangle$. In doing this, we introduce the integrals defined in the main text,
\begin{align}
\begin{split}
    &G(k)=\int_{-\infty}^\infty\phi(z)\ee^{\ii kz}~dz\\ &H(k,x)=\int_{x_0}^x\phi'(z)\ee^{-\ii kz}~dz\\ 
    &F(k) =\int_{-\infty}^\infty\phi(z)\ee^{\ii kz}H(k,z)~dz\,.
\end{split}\label{eq:integral F definition}
\end{align}
Rearranging the previous expressions in terms of these integrals, and substituting into the equations for $A$ and $B$,
\begin{align*}
A(x)&=A(x_0)\ee^{\ii k(x-x_0)}\left(1+\frac{G(k)F(k)}{1-F(k)^2}H(k,x)\right)\\
B(x)&=-A(x_0)\ee^{\ii k(x-x_0)}\left(\frac{G(k)}{1-F(k)^2}H(k,x)\right)\,,
\end{align*}
which is of the form $A(x)=t_0(k)A(x_0)\ee^{\ii k(x-x_0)}$ and $B(x)=t_1(k)A(x_0)\ee^{\ii k(x-x_0)}$, allowing us to read off the scattering coefficients stated in the main text.

\subsection{Integrals G and H and F}\label{app:Integrals G,H}

\begin{figure*}
\includegraphics[width=.8\textwidth]{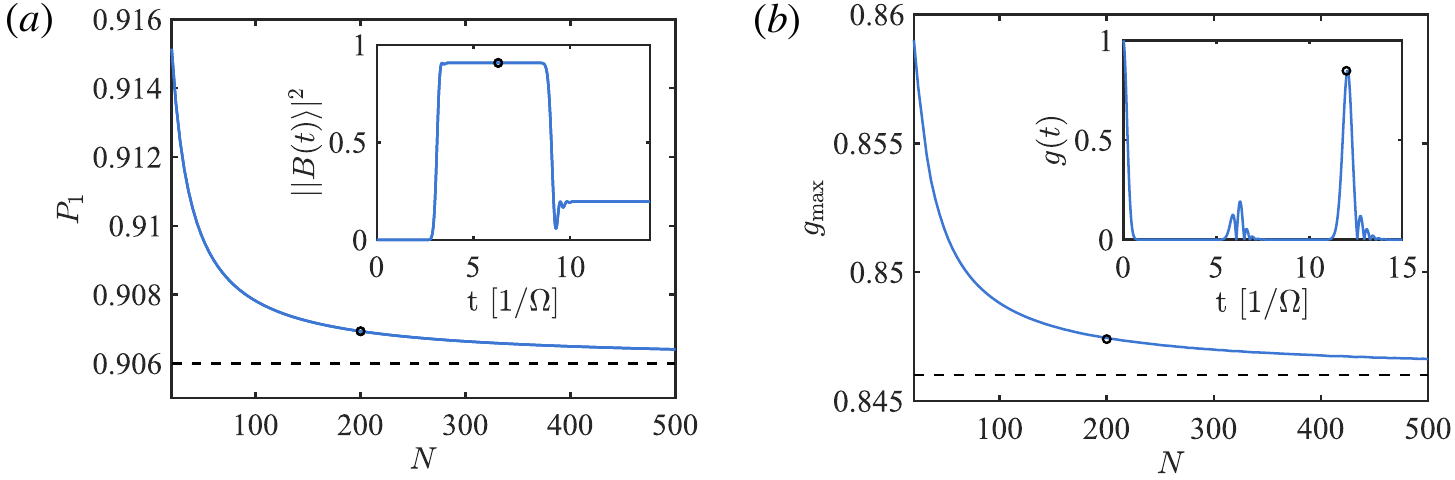}
\caption{(a) Comparison of the analytical spin coherent state scattering probability $P_1\approx0.906$ with numerical values obtained as the norm of the state on partition $B$ at $\Omega t=2\pi$ in the limit $N\to\infty$. The inset shows an example numerical calculation of $P_1$ for $N=200$, identified by the black circle. (b) Comparison of the analytical first many-body fidelity revival $g_\mathrm{max}\approx0.846$ with numerical values. The inset shows an example numerical calculation of $g_\mathrm{max}$ for $N=200$, identified by the black circle.}
\label{fig:numerics}
\end{figure*}

Given the formal solution to the scattering problem differential equations, we now seek analytical expressions for the Gaussian integrals which we use to express the scattering coefficients. We begin by computing the integral $G(k)$, since this is a Gaussian integral with infinite limits of integration, for which we know the solution well. In fact, we can see that $G(k)$ is simply the Fourier transform of $\phi(x)$ (introduced in~\eqref{eq:spin coherent wavefunctions - k} for a general spin coherent state),
\begin{align}
G(k)=2^{1/2}\pi^{1/4}\ee^{-k^2/2}
\end{align}
We note that $G(k)$ is strongly localised around $k=0$ i.e. $E=0$. This constrains the energy values with non-zero scattering amplitudes to lie within $|k|\lesssim 1$ (the standard deviation). $G(k)$ is also entirely real.

To evaluate the integral $H(k,x)$ we first note that in the large $N$ limit we can take $x_0\rightarrow -\infty$, such that 
\begin{align*}
H(k,x)&=\phi(x)\ee^{-\ii kx}+\ii k\int_{-\infty}^xdz~\phi(z)\ee^{-\ii kz}\,.
\end{align*}
Substituting the explicit expression for the vacuum state $\ket{\phi}$ as derived in~\eqref{eq:phi wavefunction}, completing the square and recognising the integral definition of the (complex) error function~\cite{abramowitz_handbook_1972} then gives
\begin{align}
\label{eq:H(k,x)}
\begin{split}
    &H(k,x)=\phi(x)\ee^{-\ii kx}+\ii k \frac{\pi^{1/4}}{2^{1/2}}\ee^{-k^2/2}\left(\mathrm{erf}(\tfrac{\ii k+x}{\sqrt{2}})+1\right).
\end{split}
\end{align}
This expression for $H(k,x)$ is used for the evaluation of $F(E)$ below. Moreover it yields the limit $H(k)=\lim_{x\to\infty}H(k,x)$ ,
\begin{align}
\label{eq:H(k)}
\begin{split}
    &H(k)=\ii k\sqrt{2}\pi^{1/4}\ee^{-k^2/2}=\ii k G(k)\,,
\end{split}
\end{align}
which is the relevant multiplicative factor for the transmission amplitudes.

We can use the form of the integral $H(k,x)$ derived in the previous section to attempt a closed-form solution of the final Gaussian integral, $F(k)$. Substituting~\eqref{eq:H(k,x)} into the definition~\eqref{eq:integral F definition}, we obtain
\begin{align*}
    F(k)
    &=1+\ii k\pi^{1/2}~\ee^{-k^2}\\
    &+\frac{\ii k}{2^{1/2}}~\ee^{-k^2/2}\int_{-\infty}^\infty\mathrm{erf}(\tfrac{\ii k+x}{\sqrt{2}})~\ee^{-x^2/2+\ii k x}dx\,.
\end{align*}
To tackle the last term in this expression, we can expand the error function in $x\ll1$ around $\ii k/\sqrt{2}$, using the formula for the $n$-th derivative of the error function,
\begin{align*}
    \mathrm{erf}^{(n)}(z)=\frac{2(-1)^{n-1}}{\pi^{1/2}}\ee^{-z^2}H_{n-1}(z)\,,
\end{align*}
for $n\geq1$ and where $H_n$ denotes the $n$-th order Hermite polynomial. This implies the power series in $x$
\begin{align*}
    \mathrm{erf}(\tfrac{\ii k+x}{\sqrt{2}})=\mathrm{erf}(\tfrac{\ii k}{\sqrt{2}})+\frac{2}{\pi^{1/2}}\sum_{n=1}^\infty \frac{(-1)^{n-1}x^n}{2^{n/2}}H_{n-1}(\tfrac{\ii k}{\sqrt{2}})
\end{align*}
The challenge of evaluating the integral $F(k)$ analytically then amounts to computing the Gaussian integrals with each of these terms. It is trivial to perform this integral for the zero-order term, which is just
\begin{align*}
    \int_{-\infty}^\infty\mathrm{erf}(\tfrac{\ii k}{\sqrt{2}})~\ee^{-x^2/2+\ii k x}dx=&\ii 2^{3/2}\mathrm{dawF}(\tfrac{k}{\sqrt{2}})
\end{align*}
where dawF denotes the extension of the Dawson function to the complex plane. For the higher-order terms in the sum in the expression, we now recall a standard Gaussian integration result,
\begin{align*}
    &\int_{-\infty}^\infty x^n\ee^{-x^2/2+\ii k x}dx=(-1)^n2^{(1-n)/2}\frac{\pi^{1/2}\ee^{-k^2/2}}{\ii^n}H_n(\tfrac{k}{\sqrt{2}})
\end{align*}
 for $n\in\mathbb{Z}^+$. This leads to the final expression for $F(k)$,
\begin{align*}
    F(k) &=1+\ii k\pi^{1/2}~\ee^{-k^2}- 2k\ee^{-k^2/2}\mathrm{dawF}(\tfrac{k}{\sqrt{2}})\\
    &-\ii2k\ee^{-k^2/2}\sum_{n=1}^\infty \frac{1}{(2\ii)^{n}n!}H_{n-1}(\tfrac{\ii k}{\sqrt{2}})H_{n}(\tfrac{k}{\sqrt{2}})\,.
\end{align*}

As explained in the main text, we are able to use these expressions to calculate the scattering probability, $P_1$, and the first maximum of the many-body fidelity, $g_\mathrm{max}$, on the infinite CBG. In Fig.~\ref{fig:numerics}, we confirm numerically that $P_1\approx0.906$ and $g_\mathrm{max}\approx0.846$ are the correct limiting values of the corresponding quantities on finite CBGs with $N$ sites as $N\to\infty$.

\bibliographystyle{apsrev4-1.bst}
\bibliography{references/Zortero.bib,references/STARTbib.bib,references/Papers3.bib,references/Bennet.bib}

\end{document}